\documentclass[preprintnumbers,amsmath,amssymb]{revtex4}
\usepackage{graphicx}
\usepackage{dcolumn}
\usepackage{bm}

\def\ket#1{\mid #1\rangle}
\def\col#1#2#3{\left(\begin{array}{c} #1 \\[0.2mm] #2 \\[0.2mm]#3\end{array}\right)}
\newcommand{\ie}{{\it i.e.}}

\begin{document}

\title{Ground state property of one-dimensional Bose-Fermi mixture}

\author{Zi-Xiang Hu, Qiu-Lan Zhang, and You-Quan Li}
\address{
Zhejiang Institute of Modern Physics, Zhejiang
University, Hangzhou 310027, P. R. China }
\date{\today}

\begin{abstract}
Bose-Fermi mixtures in one dimension are studied in detail on the basis of exact solution.
Corresponding to three possible choices of the reference state in quantum inverse scattering
method, three sets of Bethe-ansatz equations are derived explicitly.
The features of the ground state and low-lying excitations are investigated.
The ground-state phase diagram caused by the external field and chemical
potential is obtained.
\end{abstract}

\maketitle

\section{Introduction}

The study on the exactly solvable models  has been an important
topic for four decades because the perturbative approach is
unapplicable for strongly correlated systems. Particles with
$\delta$-function interaction in one dimension is a simple but
interesting model. Since Lieb and Liniger~\cite{LiebL} first
solved a spinless Bose system with $\delta$-function interaction,
there has been much more
progress~\cite{Gaudin,Woynarovich,McGuire,Gaudin67,Yang,Flicker,Sutherland,SutherlandPRB,Lai,Kulish,Korepinbook}
in this field. Particularly, when solving the two-component Fermi
system, Yang~\cite{Yang} proposed the well known Yang-Baxter
equation which has had a significant impacts in both physics and
mathematics. As the 2-component system is mostly associated with
``spin-1/2" particles that are conventionally referred as Fermi
system, the coordinate Bethe ansatz has not been applied to
2-component Bose system till recently~\cite{Li}, which is
motivated by spinor Bose-Einstein condensate in magnetically
trapped $^{87}$Rb~\cite{Williams}.

Recent observation of the superfluid to Mott insulator transition in ultracold atoms
in optical lattice~\cite{Greiner} stimulated research interests related to
strongly correlated atomic gases.
Most recently, the physics of ultracold Bose-Fermi mixtures~\cite{Lewenstein}
such as $^7$Li -$^6$Li or $^{87}$Rb -$^{40}$K mixtures become a remarkable
topics~\cite{Truscot,Schreck,Modugno,Hadzibabic02,Hadzibabic03}.
It is therefore worthwhile to investigate the features of Bose-Fermi system
on the basis of exact solutions.
The mixed system of bosons and fermions with $\delta$-function
interaction was discussed earlier in Ref.\cite{Lai} where
the ground state energy and gapless fermionic excitations are calculated
in the thermodynamic limit. However, as we aware,
the ground state phase diagram under the influence of
external field and chemical potential has not been studied yet.
These properties become more and more important nowadays due
to the rapid progress in the field of cold atomic physics.

In this paper we study one dimensional cold atomic system of
Bose-Fermi mixture systematically. Our paper is organized as
follows. In the following section we introduce the model and
derive its secular equation. In section III, we diagonalize the
secular equation by means of QISM method for three cases
respectively. Consequently, three different kinds of nested
Bethe-ansatz equations are obtained. In section IV, we explicitly
analyze the ground state and the possible low-lying excitations.
The energy-momentum spectrum for each excitation is calculated
numerically from the Bethe-ansatz equation. In the last section,
we study the system under the influence of magnetic fields and
chemical potentials for the particles to obtain the phase diagram.

\section{The model and its secular equation}\label{sec:model}

We consider a mixture of cold Bose gas and Fermi gas in one dimension. The Hamiltonian of the system is described
by Gross-Pitaevskii functional,
\begin{equation}
\mathbb{H}=\int \bigl(\sum\limits_a {\partial _x \Psi _a^* }
\partial _x \Psi _a  + c\sum\limits_{a,b} {\Psi _a^*
\Psi _a \Psi _b^* \Psi _b }\bigr) dx
\end{equation}
where the natural unit is adopted for simplicity, $c$ denotes the
interaction strength and $a, b = 1, 2, 3 $ refer to the three
components of $SU(1|2)$ fields. This is an isotropic case of the
model considered by Cazallia {et al.}\cite{AFHo03} where
approximation method was employed. We just consider this case
because the anisotropic case is unintegrable. Among these three
fields, two obey anti-commutation relation and one obeys
commutation relation. It is convenient to consider the states that
span a Hilbert space of $N$ particles
$$
\ket{\psi}=\sum_{x_1, x_2,\cdots} \psi_{a_1,\cdots a_N}(x_1,\,x_2,\cdots,x_N)
  \Psi_{a_1}(x_1)\cdots\Psi_{a_N}(x_N)\ket{0}
$$
The eigenvalue problem $\mathbb{H}\ket{\psi}=E\ket{\psi}$ becomes an
$N$-particle quantum mechanical problem with the first quantized Hamiltonian,
\begin{equation}\label{eq:Hamiltonian}
\mathcal{H}=\sum_{i=1}^{N}\frac{\partial^2}{\partial x_i^2}
 +2c\sum_{i<j}\delta(x_i-x_j)
\end{equation}
Such a system can be solved by means of Bethe-ansatz approach.
Hereby we give a brief description of the main idea of this
approach. In the domain $x_i \neq x_j$, the Hamiltonian
(\ref{eq:Hamiltonian}) reduces to that for free particles and its
eigenfunctions are therefore just superpositions of plane waves.
When two particles collide, a scattering process occurs, which is
supposed to be a pure elastic process, i.e. exchange of their
momenta. So for a given momentum $k = (k_1, k_2,\ldots, k_N)$, the
scattering momenta include all permutations of the components of
$k$. Because the Hamiltonian is invariant under the action of the
permutation group $S_N$, one can adopt the following Bethe-ansatz
wavefunction
\begin{equation}\label{eq:BAwavefunction}
\psi _a (x) = \sum\limits_{P\in S_N } {A_a (P,Q)e^{i(Pk|Qx)} }
\end{equation}
where $a = (a_1, a_2,\ldots ,a_N)$, $a_j$ denotes the $SU(1|2)$
component of the $j$th particles; $Pk$ denotes the image of a
given $k:=(k_1,k_2,\ldots, k_N )$ by a mapping $P \in S_N$;
$(Pk|Qx) = \sum\nolimits_{j = 1}^N {(Pk)_j } (Qx)_j $. And the
coefficients $A(P,Q)$ are functions of $P$ and $Q$ where $Q$
denotes the permutation such that
$0<x_{Q_1}<x_{Q_2}<\ldots<x_{Q_N}<L$. For the Bose-Fermi mixture,
the wave function should be either symmetric or antisymmetric
under permutation $\Pi^j$ depending on whether they involve Bose
label or Fermi labels.
\begin{equation}
(\Pi^j \psi )_a (x) = \pm\psi_{\Pi^j a }(x)
 \end{equation}
The $\delta$-function term in the Hamiltonian contributes a
boundary condition across the hyper-plane $x_{Q_j}=x_{Q_{j+1}}$.
Substituting the Beth-Ansatz wave function into this boundary
condition and using the continuity condition together with the
permutation symmetry, we obtain the following relation
\begin{equation}\label{eq:relation}
A_a (\Pi ^j P,Q) = {{i[(Pk)_j  - (Pk)_{j + 1} ]P^j  + c} \over
{i[(Pk)_j  - (Pk)_{j + 1} ]P^j  - c}}A_a (P,Q)
\end{equation}
where $P^j$ is the permutation between particles at $x_{Q_j}$ and
$x_{Q_{j+1}}$, which is given in appendix for concrete choice of
the Bose-Fermi labels. For example, if we consider the wave
functions of two particles, because of the different exchange
symmetries, the wave function of two bosons is
$\frac{1}{\sqrt{2}}(|\psi_1>|\psi_2>+|\psi_2>|\psi_1>)$, and
$\frac{1}{\sqrt{2}}(|\psi_1>|\psi_2>-|\psi_2>|\psi_1>)$ for two
fermions. The permutation for two bosons is $P= \left(
{\begin{array}{*{20}c}
   1 & 0 & 0 & 0  \\
   0 & 0 & 1 & 0  \\
   0 & 1 & 0 & 0  \\
   0 & 0 & 0 & 1  \\
\end{array}} \right)
$, and $-P$ for two fermions.  The matrix relating to the various
amplitudes in the same region given in Eq.(\ref{eq:relation}) is
conventionally called $S$-matrix
$$
S^{j,j+1}={{i[(Pk)_j  - (Pk)_{j + 1} ]P^j  + c} \over {i[(Pk)_j  - (Pk)_{j + 1} ]P^j  - c}}
$$
The amplitudes in region $Q$ and in its adjacent region $Q'$
are related by the R-matrix $R=P S$,
$$
A_{a_1..a_i..a_j..a_N}(Q')=(R^{ij})^{b_1...b_N}_{a_1...a_N}A_{b_1...b_N}(Q)=(R^{ij})^{b_ib_j}_{a_ia_j}A_{a_1..b_i..b_j..a_N}(Q)
$$
If $x$ is a point in the region $C(Q^{(i)})$, then $x'=(x_1,
\ldots,x_{Q_1}+L,\ldots,x_N)$ is a point in the region $C(\gamma
Q^{(i-1)})$ with $\gamma=\Pi ^{N-1}\Pi^{N-2}\cdots\Pi^2\Pi^1$.
Thus the periodic boundary condition imposes a relation between
the wave functions defined on $C(Q^{(i)})$ and $C(\gamma
Q^{(i-1)})$. Writing out this relation in terms of
Eq.(\ref{eq:BAwavefunction}), we find that the periodic boundary
conditions are guaranteed provided that $A(P;\gamma Q^{(i-1)})
e^{i{(Pk)}_1L}=A(P;Q^{(i)})$. After applying the $R$ matrix
successively, we obtain the following secular equation,
\begin{equation}\label{eq:secular}
R^{Q1,QN}  \cdots R^{Q1,Q(i + 1)} R^{Q1,Qi}  \cdots R^{Q1,Q2} A(P;Q^{(i)} )
 = e^{ - i(Pk)_1 L} A(P;Q^{(i)} )
\end{equation}

\section{Diagonalisation by Quantum Inverse Scattering Method}\label{sec:QISM}

To determine the spectrum, we should diagonalize the secular equation (\ref{eq:secular}).
This can be done by diagonalizing the operator product in the left hand side of
Eq. (\ref{eq:secular}), namely, solving the eigenvalues  of the operator
\begin{equation}
(T_j )_{a_1 a_2  \cdots a_N }^{b_1 b_2  \cdots b_N }  = (R^{jj -
1}  \cdots R^{j1} R^{jN}  \cdots R^{jj + 1} )_{a_1 a_2  \cdots a_N
}^{b_1 b_2  \cdots b_N }
\end{equation}
where
\begin{equation}\label{eq:rmatrix}
R^{ij}= \frac{(\alpha_i  - \alpha_j )I^{ij}  - ic{P}^{ij}}
{\alpha_i - \alpha_j+ic}.
\end{equation}
Since they satisfy the Yang-Baxter relation:
\begin{equation}
R^{kj} (\alpha  - \beta )R^{ki} (\alpha )R^{ji} (\beta ) = R^{ji}
(\beta )R^{ki} (\alpha )R^{kj} (\alpha  - \beta )
\end{equation}
the diagonalization can be carried out by means QISM.
For Eq. (\ref{eq:rmatrix}), a $9\times9$ monodromy can be defined in the conventional way
\begin{equation}
(T)_{a_1 a_2  \cdots a_N ,\mu}^{b_1 b_2  \cdots b_N ,\upsilon }  =
\sum\limits_{S_1 S_2  \cdots S_{N - 1} } {(R^{1A} )_{a_1
,\mu}^{b_1 ,S_1 } (R^{2A} )_{a_2 ,S_1 }^{b_2 ,S_2 }  \cdots
\cdots (R^{NA} )_{a_N ,S_{N - 1} }^{b_N ,\upsilon } }
\end{equation}
which can be written as a $3\times3$ matrix in the auxiliary space:
\begin{eqnarray}\label{eq:monotromy}
T = \left(\begin{array}{ccc}
   A & {B_1 } & {B_2 }  \\
   {C_1 } & {D_{11} } & {D_{12} }  \\
   {C_2 } & {D_{21} } & {D_{22} }  \\
  \end{array}\right)
\end{eqnarray}
in which every matrix element is an operator in quantum space.
It obeys the following RTT relations,
\begin{equation}\label{eq:rtt}
 R(\lambda  - \mu)T_1 (\lambda )T_2 (\mu) = T_2 (\mu)T_1 (\lambda
)R(\lambda  - \mu)
\end{equation}
where $T_1 = T_1  \otimes I$, $T_2 = I  \otimes T_2$
with $I$ the $3 \times 3$ unitary matrix in quantum space.

Since the Bose-Fermi mixture is $SU(1|2)$ supersymetric system,
the application of QISM becomes complicated. In the SU(3) case a
unique nested Bethe-ansatz equation was derived~\cite{LiSU3}. In
present case, however, there are three possibilities in choosing
the reference state (``pseudo-vacuum") and the successive orders
of the other states, and hence three types of nested Bethe-ansatz
equations have to be derived. In the following, we will consider
those three cases respectively.

\subsection{BFF CASE}

We first choose the Bose state as the reference state $\ket{1}$
and the other two states $\ket{2}$ and $\ket{3}$ are Fermi states,
then the permutation operator is easily written out (see $P_1$ in
(\ref{eq:permutation_bff})). This case was once noticed by
Sutherland~\cite{SutherlandPRB} in lattice model. Consequently,
the RTT relation (\ref{eq:rtt}) gives rise to 2 commutation
relations between A and B, and 8 commutation relations between B
and D. We can write them in a form of tensor product:

\begin{equation}
A(\lambda ) \otimes \bigl(B_1(\mu)\  \ B_2(\mu)\bigr) =
{\frac{k(\mu - \lambda )}
{b(\mu - \lambda )}}(B_1 (\mu)\ \ B_2 (\mu)) \otimes A(\lambda ) \\
     - {\frac{d(\mu -\lambda )} {b(\mu - \lambda )}}(B_1 (\lambda )\ \ B_2 (\lambda))
\otimes A(\mu)
\end{equation}
\begin{eqnarray}\label{eq:FCR-2}
 &&\left(\begin{array}{cc}
   {D_{11} (\lambda )} & {D_{12} (\lambda )}  \\
   {D_{21} (\lambda )} & {D_{22} (\lambda )}  \\
       \end{array}\right)
   \otimes (B_1 (\mu)\ \ B_2 (\mu))
     = (B_1 (\mu)\ \ B_2 (\mu)) \otimes
       \left(\begin{array}{cc}
       {D_{11} (\lambda )} & {D_{12} (\lambda )}  \\
       {D_{21} (\lambda )} & {D_{22} (\lambda )}  \\
            \end{array}\right)
             \times \nonumber\\
 &&\hspace{8mm}{\frac{a(\lambda - \mu)} {b(\lambda -
 \mu)}}
  \left(\begin{array}{cccc}
   1 & 0 & 0 & 0  \\[1mm]
   0 & { - \frac{d }{ a}} & {\frac{b }{ a}} & 0  \\[1mm]
   0 & {\frac{b }{ a}} & { - \frac{d }{ a}} & 0  \\[1mm]
   0 & 0 & 0 & 1
        \end{array}\right)
  - \frac{d(\lambda  - \mu)}{b(\lambda  - \mu)}(B_1 (\lambda )\ \ B_2(\lambda))\otimes
\left( \begin{array}{cc}
   {D_{11} (\mu)} & {D_{12} (\mu)}\\
   {D_{21} (\mu)} & {D_{22} (\mu)}\\
 \end{array}\right)
 \end{eqnarray}
where we used the definition:
$$
k(\alpha) = \frac{\alpha  - ic}{\alpha  + ic},\;
a(\alpha) = 1, \;
b(\alpha) = \frac{\alpha}{\alpha  + ic},\;
d(\alpha) = \frac{- ic}{\alpha + ic}
$$
It is convenient to write the R-matrix in auxiliary space in
terms of the generators of the corresponding Lie algebra (see
appendix)
\begin{equation}\label{eq:vertex}
R(\alpha) = \left( {\begin{array}{*{20}c}
   {\frac{\alpha }{{\alpha  + ic}}I - ic\frac{{2H_1  + H_2  + I}}{3}} & {\frac{{ - ic}}{{\alpha  + ic}}E_{ - \alpha _1 } } & {\frac{{ - ic}}{{\alpha  + ic}}E_{ - (\alpha _1  + \alpha _2 )} } \vspace{2mm} \\
   {\frac{{ - ic}}{{\alpha  + ic}}E_{\alpha _1 } } & {\frac{\alpha }{{\alpha  + ic}}I + \frac{{ic}}{{\alpha  + ic}}\frac{{H_2  - H_1  + I}}{3}} & {\frac{{ic}}{{\alpha  + ic}}E_{ - \alpha _2 } } \vspace{2mm} \\
   {\frac{{ - ic}}{{\alpha  + ic}}E_{(\alpha _1  + \alpha _2 )} } & {\frac{{ic}}{{\alpha  + ic}}E_{\alpha _2 } } & {\frac{\alpha }{{\alpha  + ic}}I + \frac{{ic}}{{\alpha  + ic}}\frac{{I - H_1  - 2H_2 }}{3}} \vspace{2mm} \\
\end{array}} \right)
\end{equation}
and to define a pseudo-vacuum as $N$ particles in the
highest weight state of the $SU(1|2)$ systems, \ie,
\begin{equation}\label{eq:vacuum}
\ket{\phi}=\prod\limits_{i = 1}^N \otimes\ket{1}
\end{equation}
with $\ket{1}=\col{1}{0}{0}$, $\ket{2}=\col{0}{1}{0}$ and
$\ket{3}=\col{0}{0}{1}$. We have $H_1\ket{1}=\ket{1}$,
$E_{-\alpha_1}\ket{1}=\ket{2}$,
$E_{-(\alpha_1+\alpha_2)}\ket{1}=\ket{3}$, and null if the other
operators act on the state $\ket{1}$. Thus the vertex
(\ref{eq:vertex}) becomes triangular when acting on the highest
weight state. The eigenvalues of diagonal terms of the monodromy
can be obtained as the products of the values in each quantum
space, and hence the eigenvalues of the operator A(k) and D(k)
are:
\begin{eqnarray}
 A(k)\ket{\phi}&=& \prod\limits_{l = 1}^N {\frac{{k - k_l  - ic}}{{k - k_l  + ic}}} \ket{\phi} \\
 D(k)\ket{\phi}&=& \prod\limits_{l = 1}^N {\frac{{k - k_l }}{{k - k_l  + ic}}} \ket{\phi}
\end{eqnarray}
here
\begin{equation}\label{eq:su2monodromy}
D(k) = \left( {\begin{array}{*{20}c}
   {D_{11} (k)} & {D_{12} (k)}  \\
   {D_{21} (k)} & {D_{22} (k)}  \\
\end{array}} \right)
\end{equation}
is the monodromy of SU(2) algebra nested in $SU(1|2)$.

Employing the lowering operator $B(\lambda)$ acting on the
pseudo-vacuum, we construct an eigenstate
\begin{equation}
|\omega  >  = B(\lambda _1 )B(\lambda _2 ) \cdots B(\lambda _M
)|\phi  >
\end{equation}
where $B(\lambda)$ refers to one of the $B_1(\lambda)$ or
$B_2(\lambda)$ in the monodromy (\ref{eq:monotromy}). This state
can be used to diagonalize the secular equation. The eigenvalue of
the left hand side of the secular equation (\ref{eq:secular}) can
be written as the eigenvalue of the trace of the SU$(1|2)$
monodromy matrix:
\begin{eqnarray}\label{eq:eigenSU21}
\Lambda _{SU(1|2)} (k;\lambda _1 ,\lambda _2  \cdots \lambda _M )\ket{\omega}
&=& Tr(T)|\omega  >  =\bigl( A(k) + tr( D(k) )\bigr)\ket{\omega}
   \nonumber\\
&=& \prod\limits_{l = 1}^N {\frac{{k - k_l  - ic}}{{k - k_l  +
ic}}} \prod\limits_{\alpha = 1}^M {\frac{{k(\lambda _\alpha -
k)}}{{b(\lambda _\alpha - k)}}}\ket{\omega}
    \nonumber\\
  &+& \prod\limits_{l = 1}^N
{\frac{{k - k_l }}{{k - k_l + ic}}} \prod\limits_{\alpha = 1}^M
{\frac{{a(k - \lambda _\alpha )}}{{b(k - \lambda _\alpha )}}}
\cdot \Lambda '_{SU(2)}\ket{\omega}
\end{eqnarray}

From this nested Bethe-ansatz structure we can see that there is a
SU(2) substructure in the SU$(1|2)$ system when the boson state is
chosen as the reference state. In terms of the $\check{R}$-matrix
of SU(2) appearing in Eq.(\ref{eq:FCR-2})
\begin{eqnarray}
\check{r}=\left( \begin{array}{cccc}
   1 & 0 & 0 & 0  \\
   0 & { - \frac{d }{ a}} & {\frac{b }{a}} & 0  \\[1mm]
   0 & {\frac{b }{ a}} & { - \frac{d }{ a}} & 0  \\[1mm]
   0 & 0 & 0 & 1  \\
 \end{array}\right)
 \end{eqnarray}
and the permutation matrix of SU(2)
\begin{eqnarray}
p=\left( \begin{array}{cccc}
   1 & 0 & 0 & 0  \\
   0 & 0 & 1 & 0  \\
   0 & 1 & 0 & 0  \\
   0 & 0 & 0 & 1  \\
 \end{array}\right)
\end{eqnarray}
we have the R-matrix $r=p\cdot \check{r}$. Writing it out in
auxiliary space in terms of spin operators in quantum space,
$$
r(\alpha) = \left( {\begin{array}{*{20}c}
   {\frac{{\alpha  + ic/2}}{{\alpha  + ic}}I + \frac{{ic}/2}{{\alpha  + ic}}}\sigma _z & {\frac{{ic}}{{\alpha  + ic}}}\sigma ^ -  \vspace{2mm}\\
   {\frac{{ic  }}{{\alpha  + ic}}}\sigma ^ + & {\frac{{\alpha  + ic/2}}{{\alpha  + ic}}I - \frac{ic/2} {{\alpha  + ic}}}\sigma _z  \vspace{2mm}\\
\end{array}} \right)
$$
we obtain the fundamental commutation relation from the RTT
relations, $ r\cdot \tilde{  T_1 } \cdot \tilde{T_2 } = \tilde{T_2
} \cdot \tilde{T_1 } \cdot r $, as follows
\begin{eqnarray}\label{eq:FCR-su2}
 A'(\lambda )B'(\mu ) &=& \frac{{a'(\mu  - \lambda )}}{{b'(\mu  - \lambda )}}B'(\mu )A'(\lambda )
 - \frac{{d'(\mu  - \lambda )}}{{b'(\mu  - \lambda )}}B'(\lambda )A'(\mu ) \nonumber\\
 D'(\lambda )B'(\mu ) &=& \frac{{a'(\lambda  - \mu )}}{{b'(\lambda  - \mu )}}B'(\mu )D'(\lambda )
 - \frac{{d'(\lambda  - \mu )}}{{b'(\lambda  - \mu )}}B'(\lambda )D'(\mu
 )
\end{eqnarray}
where
$$
{\rm{ a'(\alpha)=1}},\quad
{\rm{b'(\alpha) = }}\frac{{\rm{b}}}{{\rm{a}}}
= \frac{\alpha }{{\alpha  + ic}},\quad
d'(\alpha) =  - \frac{d}{a} =\frac{{ic}}{{\alpha  + ic}}
$$
and the D(k) matrix, a sub-matrix of the SU$(1|2)$ matrix, is
regarded as the SU(2) monodromy, namely
$$
\widetilde{T}(k)=\left( {\begin{array}{*{20}c}
   {A'(k)} & {B'(k)}  \\
   {C'(k)} & {D'(k)}  \\
\end{array}} \right)
$$
where $
A'(k)=D_{11}(k),B'(k)=D_{12}(k),C'(k)=D_{21}(k),D'(k)=D_{22}(k)$.

 According to the procedure of  QISM ~\cite{Korepinbook}, the
pseudo-vacuum is defined as the product of the highest weight
states of SU(2) $\ket{\phi'}=\prod\limits^M
 {\left({\begin{array}{*{20}c} 1\\[0mm] 0 \end{array}} \right)}
$
which fulfills
$$
A'(k)\ket{\phi'} = \prod\limits_{\alpha  = 1}^M {a'(k - \lambda
_\alpha  )}\ket{\phi'},\quad
D'(k)\ket{\phi'}= \prod\limits_{\alpha  = 1}^M {b'(k
- \lambda _\alpha  )}\ket{\phi'}
$$
In terms of the lowering operator $B'(k-\lambda_\alpha)$ in SU(2) monodromy,
one can construct a general state
\begin{equation}
\ket{\omega'}= B'(k - \lambda _1 )B'(k - \lambda _2 ) \cdots B'(k
- \lambda _{M'} )\ket{\phi'}
\end{equation}
Using the fundamental commutation relations (\ref{eq:FCR-su2}), one obtain that
\begin{eqnarray}
 &&\Lambda '_{SU(2)} (k;\mu_1 ,\mu_2 , \cdots \mu_{M'} )\ket{\omega'}=
 tr(\widetilde{T}(k))\ket{\omega'}
   \nonumber\\
 &&=(\prod\limits_{\alpha = 1}^M {a'(k - \lambda _\alpha )} \prod\limits_{\beta = 1}^{M'} {\frac{{a'(\mu_\beta  - k)}}{{b'(\mu_\beta - k)}}}  + \prod\limits_{\alpha = 1}^M {b'(k - \lambda _\alpha )} \prod\limits_{\beta = 1}^{M'} {\frac{{a'(k - \mu_\beta )}}{{b'(k - \mu_\beta
  )}}})\ket{\omega'}
 \nonumber\\
 &&=
 (\prod\limits_{l=\beta}^{M'}{\frac{{\mu_\beta  - k + ic}}{{\mu_\beta  - k}}}  + \prod\limits_{\alpha = 1}^M {\frac{{k - \lambda _ \alpha}}{{k - \lambda _\alpha  + ic}}} \prod\limits_{\beta = 1}^{M'} {\frac{{k - \mu_\beta  + ic}}{{k - \mu_\beta }}}
 )\ket{\omega'}
\end{eqnarray}
The unwanted terms vanishes as long as the following equations hold
\begin{equation}\label{eq:mu_lambda}
 - \prod\limits_{\beta = 1}^{M'} {\frac{{\mu_c  - \mu_\beta  - ic}}{{\mu_c  - \mu_\beta  + ic}}}  = \prod\limits_{\alpha = 1}^M {\frac{{\mu_c  - \lambda _\alpha }}{{\mu_c  - \lambda _\alpha  + ic}}}
\end{equation}
As a result, Eq. (\ref{eq:eigenSU21}) becomes
\begin{eqnarray}
 && \Lambda _{SU(1|2)}(k;\lambda _1 ,\lambda _2 , \cdots \lambda _M ) \nonumber \\
 && = \prod\limits_{l = 1}^N {\frac{{k - k_l  - ic}}{{k - k_l  + ic}}} \prod\limits_{\alpha = 1}^M {\frac{{k(\lambda _\alpha  - k)}}{{b(\lambda _\alpha  - k)}}}  + \prod\limits_{l = 1}^N {\frac{{k - k_l }}{{k - k_l  + ic}}} \prod\limits_{\alpha = 1}^M {\frac{{a(k - \lambda _\alpha )}}{{b(k - \lambda _\alpha )}}}  \cdot \Lambda '_{SU(2)}
 \nonumber\\
  && = \prod\limits_{l = 1}^N {\frac{{k - k_l  - ic}}{{k - k_l  + ic}}} \prod\limits_{\alpha = 1}^M {\frac{{\lambda _\alpha  - k - ic}}{{\lambda _\alpha  - k}}} \nonumber\\
  && + \prod\limits_{l = 1}^N {\frac{{k - k_l }}{{k - k_l  + ic}}} \prod\limits_{\alpha = 1}^M {\frac{{k - \lambda _\alpha  + ic}}{{k - \lambda _\alpha }}} (\prod\limits_{\beta = 1}^{M'} {\frac{{\mu_\beta  - k + ic}}{{\mu_\beta  - k}}}
  + \prod\limits_{\alpha = 1}^M {\frac{{k - \lambda _ \alpha}}{{k - \lambda _\alpha  + ic}}} \prod\limits_{\beta = 1}^{M'} {\frac{{k - \mu_\beta  + ic}}{{k - \mu_\beta }}}
 )\end{eqnarray}
To get rid of the unwanted terms in the expansion, the following
equations need to be satisfied
\begin{equation}\label{eq:k_lambda}
1 =  - \prod\limits_{l = 1}^N {\frac{{\lambda _\gamma   - k_l  -
ic}}{{\lambda _\gamma   - k_l }}} \prod\limits_{\beta = 1}^{M'}
{\frac{{\mu_\beta  - \lambda _\gamma  }}{{\mu_\beta  - \lambda
_\gamma + ic}}}
\end{equation} It is convenient to redefine the parameter $\lambda'_\gamma$ by $\lambda_\gamma -ic/2$. The
equations (\ref{eq:k_lambda}) and (\ref{eq:mu_lambda}) for the
complete cancellation of the unwanted terms appearing in both
procedures, together with the relation resulting from periodic
boundary conditions, $e^{-ik_\alpha  L} =\Lambda
_{SU(1|2)}(k;\lambda _1 ,\lambda _2 , \cdots \lambda _M )$ gives
rise to the Bethe-Ansatz equations
\begin{eqnarray}\label{eq:BAEBFF}
 e^{ik_j  L}  &=&  - \prod\limits_{l = 1}^N {\frac{{k_j   - k_l  + ic}}{{k_j   - k_l  - ic}}} \prod\limits_{\alpha = 1}^M {\frac{{k_j   - \lambda _\alpha  - ic/2}}{{k_j  - \lambda _\alpha  +
 ic/2}}}  \nonumber   \\
 1 &=&  - \prod\limits_{l = 1}^N {\frac{{\lambda _\gamma   - k_l  - ic/2}}{{\lambda _\gamma    - k_l  + ic/2}}} \prod\limits_{\beta = 1}^{M'} {\frac{{\lambda _\gamma-\mu_\beta  + ic/2}}{{\lambda _\gamma-\mu_\beta  - ic/2}}}  \label{aaa}\\
 1 &=&  - \prod\limits_{\alpha = 1}^M {\frac{{\mu_c  - \lambda _\alpha  - ic/2}}{{\mu_c  - \lambda _\alpha  + ic/2}}} \prod\limits_{\beta = 1}^{M'} {\frac{{\mu_c  - \mu_\beta  + ic}}{{\mu_c  - \mu_\beta  -
 ic}}}  \nonumber
\end{eqnarray}
which determine the spectrum of the SU(1$|$2) system.

\subsection{FBF CASE}

We now consider the second case, in which the Bose state is chosen
as the second state, while the first and third are Fermi state.
From the permutation operator $P_2$ (see
(\ref{eq:permutation_fbf}) ), we can get R-matrix
(\ref{eq:rmatrix}). Using the same monodromy as in Eq.
(\ref{eq:monotromy}) and the RTT relation(\ref{eq:rtt}), we get
the following communication relations:
\begin{eqnarray}\label{eq:FBF_commu1}
A(\lambda ) \otimes (B_1 (\mu)\ \ B_2 (\mu)) &=& {\frac{a(\mu -
\lambda )}
{b(\mu - \lambda )}}(B_1 (\mu)\ \ B_2 (\mu)) \otimes A(\lambda ) \nonumber\\
                                     &-& {\frac{d(\mu -\lambda )}  {b(\mu - \lambda )}}(B_1 (\lambda )\ \ B_2 (\lambda))
\otimes A(\mu) \left( \begin{array}{cc}
   1 & 0  \\
   0 & { - 1}  \\
  \end{array} \right)
\end{eqnarray}
\begin{eqnarray}\label{eq:FBF_commu2}
  \left(\begin{array}{cc}
   {D_{11} (\lambda )} & {D_{12} (\lambda )}  \\
   {D_{21} (\lambda )} & {D_{22} (\lambda )}  \\
 \end{array} \right)&\otimes&(B_1 (\mu)\ \ B_2 (\mu)) \nonumber\\
                   &=&
                        (B_1 (\mu)\ \ B_2 (\mu)) \otimes
                        \left( \begin{array}{cc}
                          {D_{11} (\lambda )} & {D_{12} (\lambda )}  \\
                          {D_{21} (\lambda )} & {D_{22} (\lambda )}  \\
 \end{array}\right){\frac{k(\lambda  - \mu)}  {b(\lambda  - \mu)}}
  \cdot\check{r}\nonumber \vspace{2mm}\\
               &-&
      {\frac{d(\lambda  - \mu)}  {b(\lambda  - \mu)}}(B_1 (\lambda )\ \ B_2
    (\lambda)) \otimes
\left( \begin{array}{cc}
   {D_{11} (\mu)} & {D_{12} (\mu)}  \\
   {D_{21} (\mu)} & {D_{22} (\mu)}  \\
 \end{array}\right)\left( \begin{array}{cc}
   1 & 0  \\
   0 & { - 1}  \\
  \end{array} \right)
 \end{eqnarray}

In this case, we can see that there is a nested SU(1$|$1)
substructure in the SU(1$|$2) system. $\check{R}$-matrix of
SU(1$|$1) appeared in Eq.(\ref{eq:FBF_commu2}) reads
\begin{eqnarray}
\check{r}=\left( \begin{array}{cccc}
   1 & 0 & 0 & 0  \\
   0 & { \frac {d }{ k}} & {\frac{b }{ k}} & 0  \\[1mm]
   0 & {\frac{b }{ k}} &   {\frac{d }{ k}} & 0  \\[1mm]
   0 & 0 & 0 & \frac{a }{ k}
 \end{array}\right)
 \end{eqnarray}
In this SU(1$|$1) substructure, the Bose state is chosen as the
highest weight state $\ket{\phi'}=\prod\limits^M
 {\left({\begin{array}{*{20}c} 1\\[0mm] 0 \end{array}} \right)}
$ when the QISM~\cite{Korepinbook} is applied.
We can obtain the Bethe-Ansatz equation by the similar procedure applied in previous case,
\begin{eqnarray}\label{eq:BAEFBF}
 e^{ik_j  L}  &=& -\prod\limits_{\alpha= 1}^M {\frac{{k_j   - \lambda _\alpha  + ic/2}}{{k_j   - \lambda _\alpha  - ic/2}}} \nonumber \\
 1 &=&  - \prod\limits_{l = 1}^N {\frac{{\lambda _\gamma   - k_l  - ic/2}}{{\lambda _\gamma   - k_l  + ic/2}}} \prod\limits_{\beta = 1}^{M'} {\frac{{\mu_\beta  - \lambda _\gamma   - ic/2}}{{\mu_\beta  - \lambda _\gamma   + ic/2}}}  \\
 1 &=&  - \prod\limits_{\alpha = 1}^M {\frac{{\mu_c  - \lambda _\alpha  + ic/2}}{{\mu_c  - \lambda _\alpha  - ic/2}}}
 \nonumber
 \end{eqnarray}

\subsection{FFB CASE}

We turn to the case that the Bose state is chosen as the
third state, and one of the Fermi state as the reference state,
the other Fermi state as the second state. In terms of the permutation
matrix $P_3$ (see(\ref{eq:permutation_ffb}) ), we can get the
R-matrix(\ref{eq:rmatrix}). The RTT relation(\ref{eq:rtt}) gives
rise to the following communication relations:
\begin{eqnarray}\label{eq:FFB_commu1}
A(\lambda ) \otimes (B_1 (\mu)\ \ B_2 (\mu)) &=& {\frac{a(\mu -
\lambda )}
{b(\mu - \lambda )}}(B_1 (\mu)\ \ B_2 (\mu)) \otimes A(\lambda ) \nonumber\\
                                     &-& {\frac{d(\mu -\lambda )}  {b(\mu - \lambda )}}(B_1 (\lambda )\ \ B_2 (\lambda))
\otimes A(\mu) \left( \begin{array}{cc}
   1 & 0  \\
   0 & { - 1}  \\
  \end{array} \right)
\end{eqnarray}

\begin{eqnarray}\label{eq:FFB_commu2}
  \left(\begin{array}{cc}
   {D_{11} (\lambda )} & {D_{12} (\lambda )}  \\
   {D_{21} (\lambda )} & {D_{22} (\lambda )}  \\
 \end{array} \right)&\otimes&(B_1 (\mu)\ \ B_2 (\mu)) \nonumber\\
                   &=&
                        (B_1 (\mu)\ \ B_2 (\mu)) \otimes
                        \left( \begin{array}{cc}
                          {D_{11} (\lambda )} & {D_{12} (\lambda )} \\
                          {D_{21} (\lambda )} & {D_{22} (\lambda )} \\
 \end{array}\right){\frac{a(\lambda  - \mu)}  {b(\lambda  - \mu)}}
  \left( \begin{array}{cccc}
   1 & 0 & 0 & 0  \\[1mm]
   0 & {  \frac{d }{ a}} & {\frac{b }{ a}} & 0  \\[1mm]
   0 & {\frac{b }{ a}} & {  \frac{d }{ a}} & 0  \\[1mm]
   0 & 0 & 0 & \frac{k }{ a}
 \end{array}\right)\nonumber\\
               &-&
    {\frac{d(\lambda  - \mu)}{b(\lambda  - \mu)}}(B_1 (\lambda )\ \ B_2
    (\lambda)) \otimes
\left( \begin{array}{cc}
   {D_{11} (\mu)} & {D_{12} (\mu)}  \\
   {D_{21} (\mu)} & {D_{22} (\mu)}  \\
 \end{array}\right)\left( \begin{array}{cc}
   {-1} & 0  \\
   0 & 1  \\
  \end{array} \right)
 \end{eqnarray}
They are almost the same as that in the ``FBF" case, there is also
a nested SU(1$|$1) substructure whose $\check{r}$ matrix appears
in Eq.(\ref{eq:FFB_commu2}). The following nested Bethe-ansatz
equation is obtained
\begin{eqnarray}\label{eq:BAEFFB}
 e^{ik_j  L}  &=& -\prod\limits_{\alpha = 1}^M {\frac{{k_j   - \lambda _\alpha  + ic/2}}{{k_j   - \lambda _\alpha  - ic/2}}}  \nonumber\\
 1 &=&  - \prod\limits_{l = 1}^N {\frac{{\lambda _\gamma   - k_l  - ic/2}}{{\lambda _\gamma   - k_l  + ic/2}}\prod\limits_{\alpha= 1}^M {\frac{{\lambda _\gamma   - \lambda _\alpha  + ic}}{{\lambda _\gamma   - \lambda _\alpha  - ic}}} } \prod\limits_{\beta = 1}^{M'} {\frac{{\mu_\beta  - \lambda _\gamma   + ic/2}}{{\mu_\beta  - \lambda _\gamma   - ic/2}}}  \\
 1 &=&  - \prod\limits_{\alpha = 1}^M {\frac{{\mu_c  - \lambda _\alpha  - ic/2}}{{\mu_c  - \lambda _\alpha  + ic/2}}}
\nonumber
 \end{eqnarray}
which were also derived in Ref.~\cite{Lai} by means of the
coordinate Bethe ansatz.

\section{Ground state and its low-lying excitations}\label{sec:groundState}

\subsection{BFF CASE}
In this case, the boson state is chosen as the reference state.
there are $N$ particles in all. After $M$ lower-operators B act on
the reference state, there just $N-M$ bosons. Analogously, after
$M'$ lower-operators act on the second state, there are $M-M'$
fermions of species 1 and $M'$ fermions of species 2. The same
analysis in other two case. Taking the logarithm of
(\ref{eq:BAEBFF}), we have
\begin{eqnarray}
k_j L&=& 2 \pi I_j + \sum_{l=1}^N \Theta_1 (k_j-k_l) +
\sum_{\alpha=1}^M \Theta_{-1/2}(k_j-\lambda_\alpha)   \nonumber  \\
2 \pi J_\gamma &=& \sum_{\l=1}^N \Theta_{-1/2}(\lambda_\gamma-k_l)+ \sum_{\beta=1}^{M'}{\Theta_{1/2}(\lambda _\gamma-\mu_\beta)}   \label{bbb} \\
2 \pi J_c '&=& \sum_{\alpha=1}^M
\Theta_{-1/2}(\mu_c-\lambda_\alpha)+\sum_{\beta=1}^{M'}\Theta_1(\mu_c-\mu_\beta)
\nonumber
\end{eqnarray}
where $\Theta_n (x)=-2 \tan^{-1}({x / nc})$ and $I_j$ is an
integer (half-odd integer) if $N-M-1$ is even (odd), while
$J_\gamma$ is an integer (half-odd integer) if $N-M'-1$ is even
(odd), and $J_c '$ is an integer (half-odd integer) if $M-M'-1$ is
even (odd). Once all roots \{$k_j$, $\lambda_\gamma$, $\mu_c$\}
are solved from the above equations (\ref{bbb}) for a given set of
quantum numbers \{$I_j$, $J_\gamma$, $J'_c$\}, the energy and the
momentum will be calculated by
\begin{equation}\label{ccc}
 E = \sum_{j=1}^N {k_j^2}, P =\frac{2\pi }{ L} \left[\sum_{j=1}^N {I_j}
-\sum_{\gamma=1}^M{J_\gamma} -\sum_{c=1}^{M'} {J_{c}'}\right].
\end{equation}

\subsubsection{The ground state}

It is easy to show that the right-hand side of the first equation
of equations(\ref{bbb}) is a monotonically increasing function of
$k_j$, i.e. if $I_i<I_j$, then $k_i<k_j$. Thus the configuration
of $\{I_j\}$ for the ground state is given by successive integers
of half-integers symmetrically arranged around zero. Given a set
of quantum numbers $I_j,J_\gamma,J'_c$ with the solutions
${k_j,\lambda_\gamma,\mu_c}$, it is useful to consider the
weak-coupling limit $c\rightarrow 0^+$. Due to $\Theta_{\pm
n}(x)\rightarrow \mp \pi {\rm sgn}(x)$, Eqs.(\ref{bbb}) become
\begin{eqnarray}\label{ddd}
2 \pi I_j &=& k_j L + \sum_{l=1}^N \pi {\rm
sgn}(k_j-k_l)-\sum_{\alpha=1}^M \pi {\rm sgn}(k_j-\lambda_\alpha)
\nonumber  \\
 2 \pi J_\gamma &=& \sum_{l=1}^N \pi {\rm sgn}(\lambda_\gamma-k_l)-\sum_{\beta=1}^{M'} \pi {\rm sgn}(\lambda_\gamma-\mu_\beta)  \\
2 \pi J_c' &=& \sum_{\alpha=1}^M \pi {\rm
sgn}(\mu_c-\lambda_\alpha)-\sum_{\beta=1}^{M'} \pi {\rm
sgn}(\mu_c-\mu_\beta) \nonumber
\end{eqnarray}
The subscripts of the rapidities $k_j$, $\lambda_\gamma$, $\mu_c$
are chosen in such a way that their quantum numbers
${k_j,\lambda_\gamma,\mu_c}$ are all ranged in an increasing
order. Then we have
\begin{eqnarray}\label{eee}
2(I_{j+1}-I_j-1) &=& {(k_{j+1}-k_j)\frac{L} { \pi}} -\sum_{\alpha=1}^M \left[ {\rm sgn}(k_{j+1}-\lambda_\alpha)-{\rm sgn}(k_j-\lambda_\alpha) \right]  \nonumber \\
2(J_{\gamma+1}-J_\gamma) &=& \sum_{l=1}^N \left[ {\rm
sgn}(\lambda_{\gamma+1}-k_l)-{\rm sgn}(\lambda_\gamma-k_l) \right]
-\sum_{\beta=1}^{M'}\left[ {\rm
sgn}(\lambda_{\gamma+1}-\mu_\beta)-{\rm
sgn}(\lambda_\gamma-\mu_\beta)\right]    \\
2(J_{c+1}'-J_c'+1) &=& \sum_{\alpha=1}^M \left[ {\rm
sgn}(\mu_{c+1}-\lambda_\alpha)-{\rm sgn}(\mu_c-\lambda_\alpha)
\right]\nonumber
\end{eqnarray}
Thus, if $J_{c+1}'-J_c'= m$, there will be $m+1$ solutions of
$\lambda_\alpha$ between $\mu_c$ and $\mu_{c+1}$, and if
$I_{j+1}-I_j = n$, correspondingly with a $\lambda_\alpha$
satisfying $k_j < \lambda_\alpha < k_{j+1}$, then we will get
$k_{j+1}-k_j=\frac{ 2 \pi n }{ L}$, comparing with no such a
$\lambda_\alpha$ there will be $k_{j+1}-k_j =\frac{ 2 \pi (n-1) }{
L}$. Obviously, such a $\lambda_\alpha$ always repels the $k$
rapidity away, then leading to the rising of the energy. Thus the
ground state of this system should be with no $\lambda_\alpha$
lying in $k_j$.

In the strong coupling limit $c \rightarrow \infty$, we have
$\tan^{-1}\frac{x }{ c}\simeq{\frac{x} { c}}$. Substituting these
to the secular equations (\ref{bbb}) for the ground state ($M=0$
,$M'=0$) and the low-lying excited state with ($M=1$,$M'=0$), the
secular equations become:
\begin{eqnarray}\label{fff}
k_j L &=& 2\pi I_j  - 2\sum\limits_{l = 1}^N {\frac{k_j  - k_l }
 {c}} \nonumber \\
k_j^{'} L &=& 2\pi I_j^{'}  - 2\sum\limits_{l = 1}^N
{\frac{k_j^{'} - k_l } { c}}  + 2{\frac{k_j^{'}  - \lambda _1 }
{c/2}} \nonumber
\end{eqnarray}
Here we change the ground state by adding one $\lambda_\alpha$
which leads to $I_j-I_j^{'}=1/2$. From the two equations above we
can get:
\begin{eqnarray}
(k_{j + 1}  - k_j )L[1 + {\frac{2N} {c L}}] = 2\pi \nonumber
\\
(k_{j + 1}^{'} - k_j^{'} )L [1 + {\frac{2N - 4}  {c L}}] = 2\pi
\label{ggg}
\end{eqnarray}

In order to analyzing the low-lying excited characters of the
system more conveniently, we introduce density of roots
$$
\rho (k_j ) = \frac{1 }{ {L(k_{j + 1}  - k_j )}}\   \   , \ \
\sigma (\lambda _\gamma  ) = \frac{1 }{ {L(\lambda _{\gamma + 1} -
\lambda _\gamma )}}\ \  , \ \ \omega (\mu _c ) = \frac{1 }{L(\mu
_{c + 1} - \mu _c )}
$$
In thermodynamics limit, we have $ \rho (k) = \frac{1 }{ L}{\frac{dI(k)} {dk}}$, corresponding to $\sigma
(\lambda) = \frac{1 }{ L}{\frac{dJ(\lambda )}{d\lambda }}$ and $\omega(\mu)=\frac{1 }{ L}{\frac{dJ'(\mu)} {d
\mu}}$. In terms of these densities, the energy and the momentum per length are given by
\begin{equation}
E/L = \int {k^2 \rho (k)dk}, \quad  P/L = \int {k\rho (k)dk}
\end{equation}
while N, M and M' are determined by
\begin{equation}
N/L = \int {\rho (k)dk} \ \ , \ \
 M/L = \int {\sigma (\lambda )d\lambda}\ \ , \ \ M'/L = \int {\omega (\mu )d\mu }
\end{equation}
\begin{figure}
\includegraphics[width=5cm,height=4.6cm]{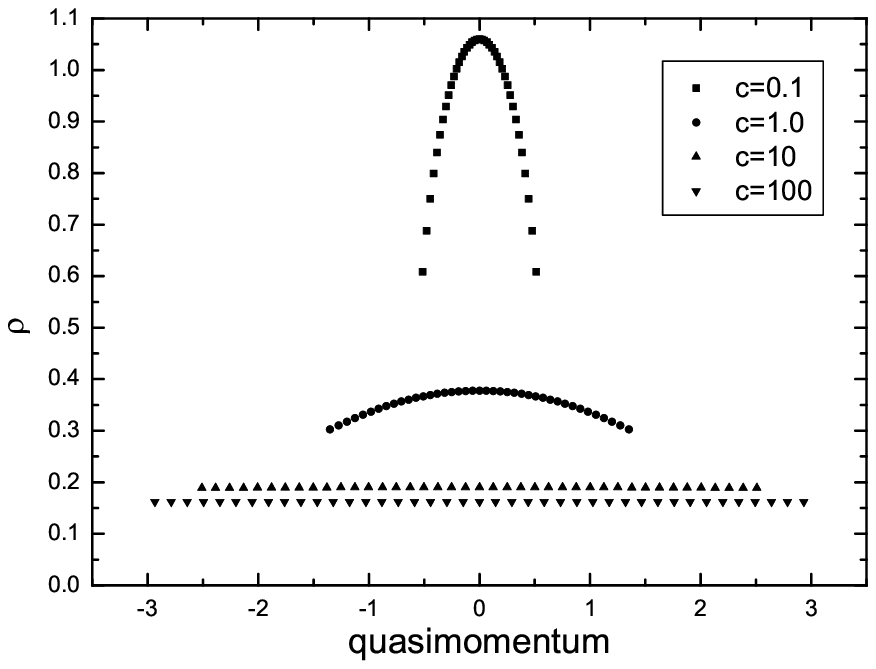}
\includegraphics[width=5cm,height=4.6cm]{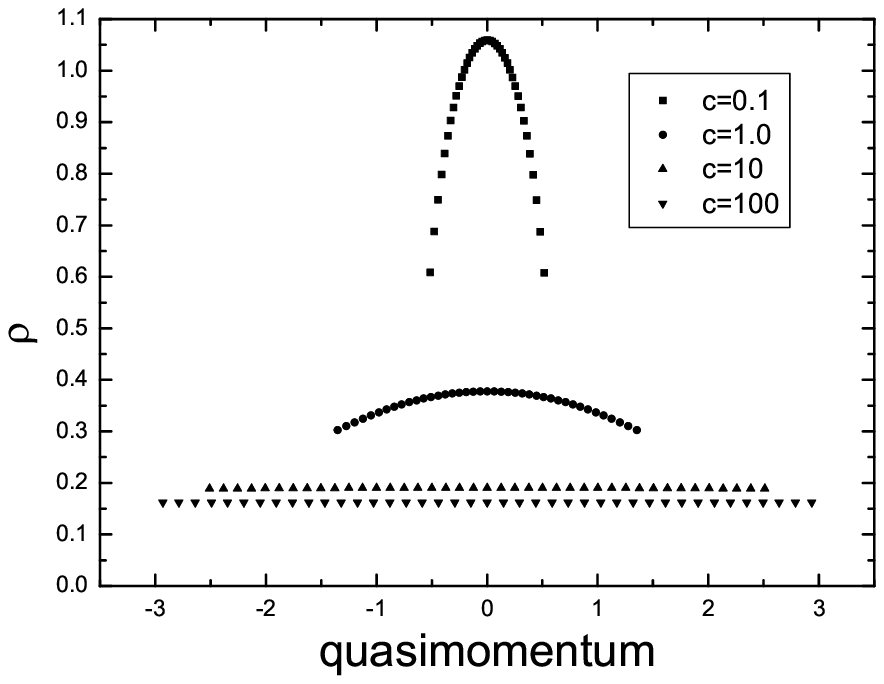}
\includegraphics[width=5cm,height=4.6cm]{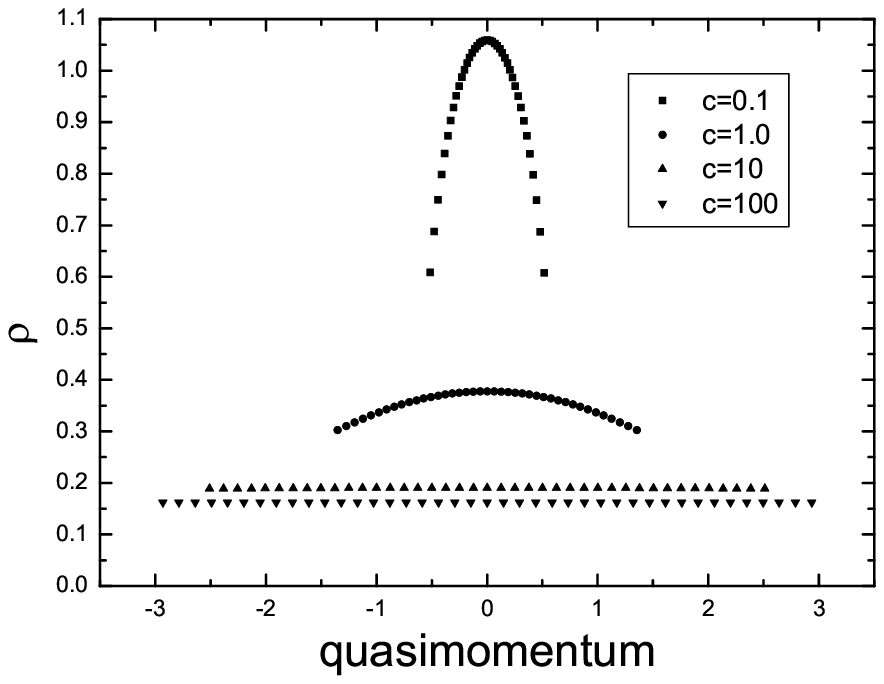}
\caption{\label{Figure Graph1} The density of state in $k$-space
for the ground state(left), adding one fermion(middle), and adding
three fermions(right). The distribution changes from a histogram
to a narrow peak gradually for the coupling from strong to weak.
The figure is plotted for $c=100, 10, 1, 0.1$.}
\end{figure}
where $ K_n (x) = \frac{1 }{ \pi }{\frac{nc/2}  {n^2 c^2 /4 +
x^2 }}$ the density of the state
satisfies the integral equation
\begin{equation}
\rho _0 (k) =\frac {1 }{ {2\pi }} + \int_{ - k_F }^{k_F } {K_2 (k
- k')} \rho _0 (k')dk'
\end{equation}
in the thermodynamic limit, where $\rho_0(k)$ and $k_F$ are the
density and integration limit for the ground state, respectively.
We solved the secular equation for 42 particles with $M=M'=0$
numerically, and the density of the ground state is depicted in
Fig.\ref{Figure Graph1}(left) for different coupling constants.

Comparing with the ground state, we plot the spectrum for the
low-lying excitation $(M=1,M'=0)$ in Fig.\ref{Figure
Graph1}(middle). The density of state are slightly compressed
compared with Fig.\ref{Figure Graph1}(middle). It is not obvious
in numerical results in Fig.\ref{Figure Graph1}(middle) comparing
to Fig.\ref{Figure Graph1}(left), so we increase $M$ from 1 to 3,
and the curves are depressed more evidently in the system with
$N=42$ (Fig.\ref{Figure Graph1}(right)). As the value of $M$
rises, the number of fermions rises correspondingly, and the
larger the number of the fermions is, the higher the energy should
be. As a result, the ground state contains only bosons which
agrees with the results of our asymptotic analysis.

\subsubsection{Particle-hole excitation}

\begin{figure}
\includegraphics[width=10.6cm,height=4.6cm]{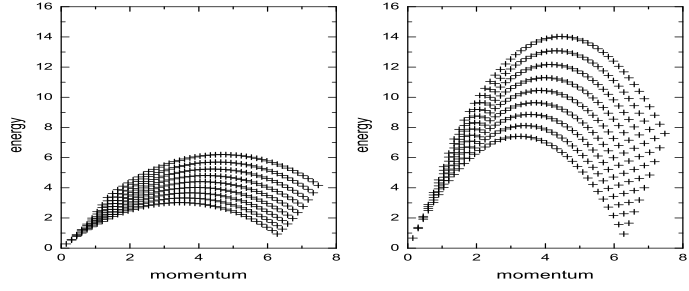}
\caption{\label{Figure Graph4}
Particle-hole excitation for $c=1.0$ (left) and $c=10.0$(right).}
\end{figure}

The quantum numbers for the ground state in N particles system are
$ \{ I_j \} = \{  - (N - 1)/2, \cdots (N - 1)/2\} , \{ J_\gamma \}
=\{ J'_c \}={\it empty}$. If we add a hole to the ground state,
then the quantum numbers take the values $ I_1 =  - (N - 1)/2 +
\delta _{1,j}$ for $(1\leq j_1\leq N)$,
$I_j=I_{j-1}+\delta_{j,j_1}$ for $(j=2,...,N-1)$, $I_N=I_0,
(I_0>(N-1)/2)$, we call it the particle-hole excitation. In Fig.
\ref{Figure Graph4} the excitation spectrum is plotted with
coupling numbers ($c=1.0,\, 10.0$).

In the thermodynamic limit, we use the expression $\rho (k) = \rho _0 (k) + \rho _1 (k)/L
$, then removing one $I$ from the original symmetric sequence and
adding a new $I_n$ outside it, we have
\begin{equation}
\rho _1 (k) + \delta (k - \overline k ) = \int {K_2 (k - \overline
k )} \rho _1 (k')dk' + K_2 (k - k_p )
\end{equation}
The excited energy consists of two terms
$ \Delta E = \frac{1 }{ L}\int {\rho _1 (k)k^2 dk}  + \frac{1 }{ L}k_p^2
= \xi _h (\overline k ) + \xi _a (k_p )$,
where $\xi_h$ is holon's energy and $\xi_a(k_p)$ is particle's energy,
and they can be calculated by
\begin{eqnarray}
\rho _1^h (k,\overline k ) &=& \int {K_2 (k - k')} \rho _1^h (k' -
\overline k )dk' - K_2 (k - \overline k )  \nonumber \\
 \xi _h (\overline k ) &=&
- {\overline {k}}^2  + \int_{ - k_F }^{k_F } {k^2 \rho _1^h
(k,\overline k )dk}  \label{hhh}
\end{eqnarray}

\subsubsection{Add one fermion}
If we add one fermion into the ground state, this excitation can be
characterized by moving the quantum number $J_1$ in the following
region:
$$
 - (N - 1)/2 < J_1  < (N - 1)/2
$$
\begin{figure}
\includegraphics[width=10.6cm,height=4.6cm]{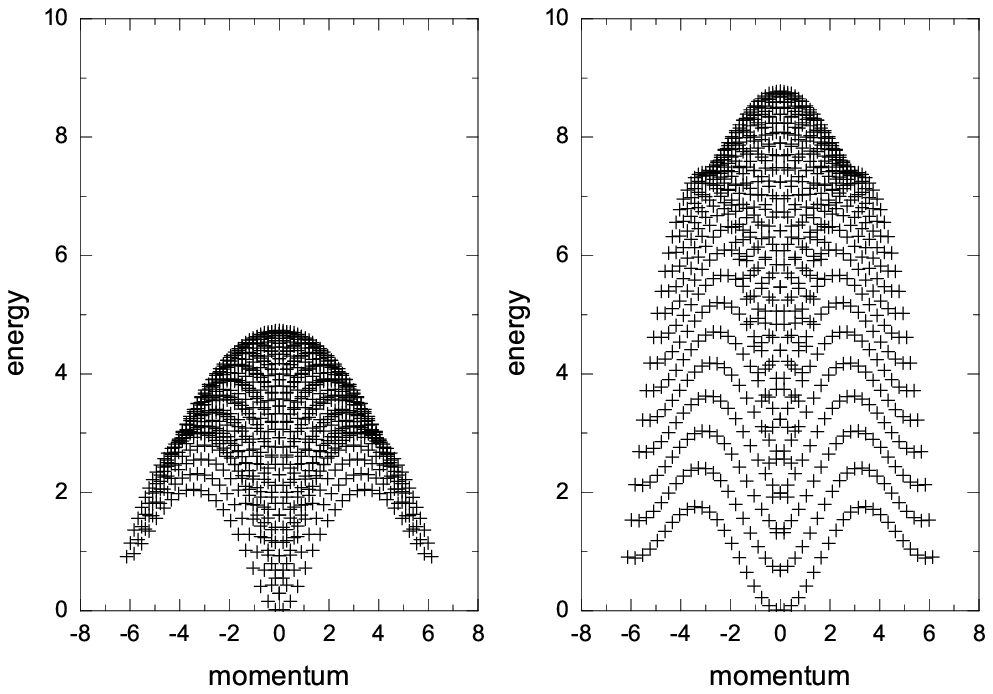}
\includegraphics[width=10.6cm,height=4.6cm]{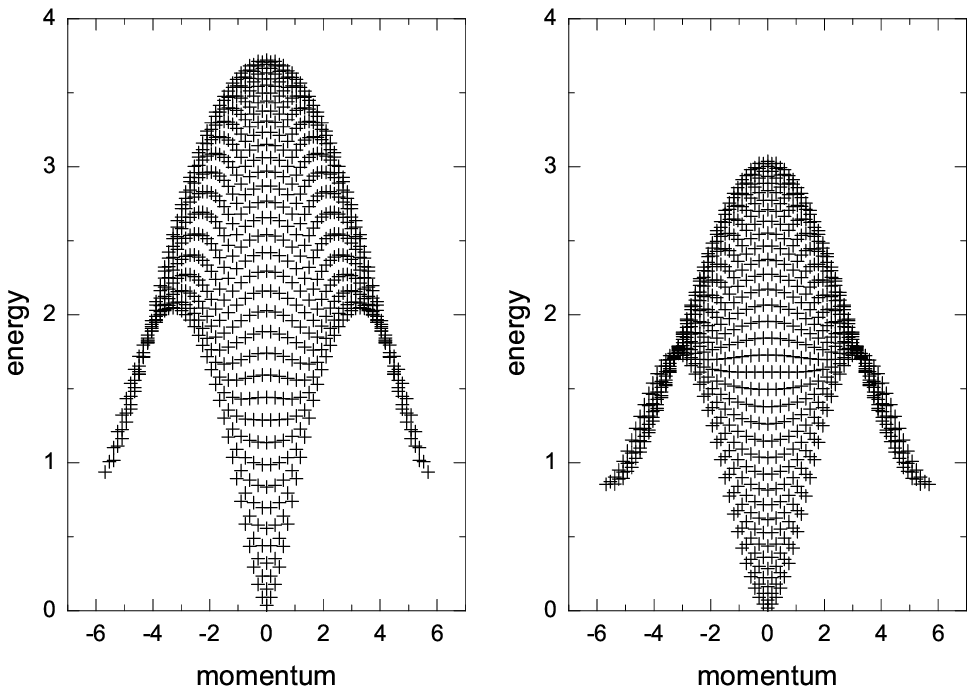}
\caption{\label{Figure Graph5}
Top panels show the one-fermion excitation for
$c=1.0$ (left) and $c=10.0$(right);
Bottom panels show the two-fermion excitation for
$c=1.0$ (left) and $c=10.0$(right).}
\end{figure}
We describe this phenomenon in Fig.\ref{Figure Graph5}(a).
Replacing one boson by one fermion corresponds to a two-parameter
excitation. Its energy is given by $\Delta E = \int {k^2 \rho _1^c
(k,\lambda )dk}$ with $\rho_1(k)$ solving
\begin{equation}
\rho _1 (k) + \delta (k - \overline k ) = \int {K_2 (k - k')} \rho
_1 (k')dk' + K_1 (k - \lambda _1 )
\end{equation}
Then we have one fermion excitation $ \Delta E = \xi _h (\overline k ) + \xi _c (\lambda )$, where $\xi _h
(\overline k )$ is the same to Eqs.(\ref{hhh}) and $\xi_c$ is defined by $ \xi _c (\lambda ) = \int {k^2 \rho _1^c
(k,\lambda )dk}$ with
\begin{equation}
\rho _1^c (k,\lambda ) = \int_{ - k_F }^{k_F } {K_2 (k - k_l )\rho
_1^c (k_l ,\lambda )dk_l }  - K_1 (k - \lambda _1 )
\end{equation}

\subsubsection{Two fermions excitation}

If two spin up fermions or two spin down fermions are permitted in
this system, then the arrangement of their quantum numbers is
$$
 - (N - 1)/2 \le J_1 \le J_2 \le (N - 1)/2
$$

Numerical calculation for this type of excitation is shown in
Fig.\ref{Figure Graph5}(b). Furthermore, if we introduce one
spin-up and one spin-down fermions in this system, the results are
the same as Fig.\ref{Figure Graph5}(b). Comparing to the system of
pure bosons, we found that such excitation likes
isospinon-isospionon excitation in two-band $SU(2)$
system\cite{Zhang}.

\subsection{FBF CASE}

From Eqs.(\ref{eq:BAEFBF}), we know there are $N-M$ fermions of
species $1$, $M-M'$ bosons and $M'$ fermions of species $2$.
Taking the logarithm of Eqs.(\ref{eq:BAEFBF}), we get
\begin{eqnarray}
k_j L &=& 2 \pi I_j + \sum_{\alpha=1}^M
\Theta_{1/2}(k_j-\lambda_\alpha)   \nonumber \\
2 \pi J_\gamma &=& \sum_{\l=1}^N \Theta_{-1/2}(\lambda_\gamma-k_l)+ \sum_{\beta=1}^{M'}{\Theta_{1/2}(\lambda _\gamma-\mu_\beta)}   \label{sub} \\
2 \pi J_c' &=& \sum_{\alpha=1}^M
\Theta_{-1/2}(\mu_c-\lambda_\alpha) \nonumber
\end{eqnarray}
The quantum number $I_j$ take integer or half-integer values, depends on whether $M$ is even or odd. And $J_c'$
take half-integer(integer) values when $M$ is even(odd). While $J_\gamma$ is integer (half-integer), if $N-M'-1$
is even (odd). In the weak-coupling limit, $c\rightarrow 0^+ $, $\Theta_{1/2}(x)\rightarrow - \pi sgn(x)$, and
$\Theta_{-1/2}(x)\rightarrow  \pi sgn(x)$ for $x>>1$, hence Eqs.(\ref{sub}) becomes
\begin{eqnarray}\label{editor}
2 I_j &=& \frac{ k_j L }{ \pi} + \sum_{\alpha=1}^M {\rm
sgn}(k_j-\lambda_\alpha)
\nonumber  \\
 2 J_\gamma &=& \sum_{l=1}^N {\rm sgn}(\lambda_\gamma-k_l)-\sum_{\beta=1}^{M'}  {\rm sgn}(\lambda_\gamma-\mu_\beta)  \\
2 J_c' &=& \sum_{\alpha=1}^M \pi {\rm
sgn}(\mu_c-\lambda_\alpha)-\sum_{\beta=1}^{M'}  {\rm
sgn}(\mu_c-\mu_\beta) \nonumber
\end{eqnarray}
As choosing $I_j, J_\gamma, J_c '$ in an increasing order, for a given $M$ and $M \leq N$ with the rules of Young
tableau, the minimum value of the left-hand side of the third equation of Eqs. (\ref{editor}) is $-M+2$.
Therefore, the smallest $\lambda_\alpha$ must be smaller than the smallest $\mu_c$. Otherwise the left-hand side
would be $-N$, and if we take the maximum value of the left-hand side $M-2$, correspondingly the largest
$\lambda_\alpha$ must be larger than the larger $\mu_c$. In other words, the presence of $\mu_c$ is only allowed
in $\lambda$ space. Furthermore, we can obtain $\left | J_c  \right | \le (N-M')/2$, when there is no spin down
fermion ($M'=0$), all $J_\gamma$ like to stay in the $I_j$ sequence.

Supposing the rapidities $k_j$, $\lambda_\gamma$, $\mu_\beta$ are
in an increasing order as $I_j$, $J_\gamma$, $J_c'$, then we have
\begin{eqnarray}\label{sud}
2(I_{j+1}-I_j) &=& \frac{(k_{j+1}-k_j)L }{\pi} -\sum_{\alpha=1}^M \left[ {\rm sgn}(k_{j+1}-\lambda_\alpha)-{\rm sgn}(k_j-\lambda_\alpha) \right]  \nonumber \\
2(J_{\gamma+1}-J_\gamma) &=& \sum_{l=1}^N \left[ {\rm
sgn}(\lambda_{\gamma+1}-k_l)-{\rm sgn}(\lambda_\gamma-k_l) \right]
-\sum_{\beta=1}^{M'}\left[ {\rm
sgn}(\lambda_{\gamma+1}-\mu_\beta)-{\rm
sgn}(\lambda_\gamma-\mu_\beta)\right]  \nonumber  \\
2(J_{c+1}'-J_c') &=& \sum_{\alpha=1}^M \left[ {\rm
sgn}(\mu_{c+1}-\lambda_\alpha)-{\rm sgn}(\mu_c-\lambda_\alpha)
\right]
\end{eqnarray}
Therefore, let $I_{j+1}-I_j = n$ and there will be $k_{j+1}-k_j =
2 \pi(n-1) / L$ existing a $\lambda_\alpha$ between $k_j$ and
$k_{j+1}$. Otherwise $k_{j+1}-k_j = 2 \pi n / L$. That means
existing a $\lambda_\alpha$ in $k_j$ space will decrease the
system's energy, so the ground state should have more
$\lambda_\alpha$. If $J_{c+1}'-J_c' = m$, we know there will be
$m$ solutions $\lambda_\alpha$ satisfying
$\mu_c<\lambda_\alpha<\mu_{c+1}$. When $J_{\gamma+1}-J_\gamma=
m'$, $m'$ is the integer which equals or surpasses one, then there
must be some $k_j$ lying in neighboring $\lambda_\gamma$, and if
we have the larger the number of $\mu_\beta$ between
$\lambda_\gamma$ and $\lambda_{\gamma+1}$, the larger the number
of $k_j$ we hope. If $m'$ is large enough, we wish that $k_j$ will
be large enough, and
 $\mu_\beta$ will be small enough. When $m'$ equals one, there is
 only one $k_j$ between $\lambda_\gamma$ and $\lambda_{\gamma+1}$,
 this state contains no $\mu_\beta$. That is to say $k$ and
 $\lambda$ alternate.

Then in the strong-coupling limit $c\rightarrow \infty$, $\tan^{-1}(x/c)\rightarrow x/c $ and Eqs.(\ref{sub}) give
rise to
$$
k_j L = 2\pi I_j - 2M \frac{k_j - \lambda _\alpha  }{c/2}
$$
Furthermore, the new form will be
\begin{equation}
\left( k_{ j + 1}  - k_j  \right)L\left[ {1 + {\frac{4M}
{cL}}} \right] = 2\pi   \\
\end{equation} From the formula above, we know if $M$ approaches to $N$, $\Delta k=k_{j+1}-k_j$ is the
smallest, so does the energy. Therefore, the $M=N$ state is the
ground state.

We describe the density of ground state by numerical approaches,
and find that is the same as Fig.\ref{Figure Graph1}(left). If we
permit $M=N-3$, i.e. there are three spin up fermions lying in the
ground state, the density of state is also shown in
Fig.\ref{Figure Graph1}(right). It is obvious that the more the
fermions lying in this system, the higher the energy will be. Thus
the ground state is exactly all bosons without one fermion, which
verifies the analysis in case 1. Furthermore we remove one of the
$I'$s from the ground state sequence and add a ``new" $I_0$
outside. Excited states are obtained by varying the quantum number
as
$$
{I_j} = {-(N - 1)/2,\cdots,-(N - 1)/2+i-1,-(N - 1)/2+i+1,\cdots,(N - 1)/2,I_0},
$$
$$
 {J_\gamma} = {-(N -1)/2,\cdots,(N - 1)/2},
$$
consequently we get the excitation spectrum which isn't different from Fig.\ref{Figure Graph4}.

For this case the excitation of adding one fermion is obtained
from $M=N-1$, which contains one free parameter in the
$I$-sequence and one free parameter in the $J$-sequence. Thus the
order of quantum numbers is:
\begin{eqnarray}
I_1 &=& -N/2+1+\delta_{j_1,1} \nonumber\\
I_j  &=& I_{j-1}+1+\delta_{j_1,j} \ \ \ \  \  (j=2,\cdots,N) \nonumber\\
J_1  &=& -N/2+\delta_{\alpha_1,1},\nonumber\\
J_\alpha &=& J_{\alpha-1}+1+\delta_{\alpha,\alpha_1} \ \ \ \ (\alpha=2,\cdots,M) \nonumber
\end{eqnarray}
where $1\leq j_1\leq N+1, 1\leq \alpha_1 \leq M+1$. The excitation
spectrum is shown in Fig.\ref{Figure Graph5}(a), which is also
consistent with the excitation of adding one fermion in case $1$.

Comparing to the case 1, when we add two spin up fermions
($M=N-2$), there are two free parameters in the $J$-sequence. The
result is depicted in Fig.\ref{Figure Graph5}(b).

\subsection{FFB CASE}

There are $N-M$ fermions of species $1$, $M-M'$ fermions of
species $2$ and $M'$ bosons in Eqs.(\ref{eq:BAEFFB}). Also taking
the logarithm of these equations we obtain
\begin{eqnarray}\label{equationbb}
k_j L &=& 2 \pi I_j + \sum_{\alpha=1}^M
\Theta_{1/2}(k_j-\lambda_\alpha)   \nonumber \\
2 \pi J_\gamma &=& \sum_{l=1}^N \Theta_{-1/2}(\lambda_\gamma-k_l)
+ \sum_{\alpha=1}^M \Theta_1 (\lambda_\gamma-\lambda_\alpha)+
\sum_{\beta=1}^{M'} \Theta_{-1/2}(\lambda_\gamma-\mu_\beta)\\
2 \pi J_c' &=& \sum_{\alpha=1}^M
\Theta_{-1/2}(\mu_c-\lambda_\alpha) \nonumber
\end{eqnarray}
where $I_j$ is an integer (half-odd integer) if $N-M-1$ is even
(odd), $J_\gamma$ is an integer (half-odd integer) if $N-M-M'-1$
is even (odd), and $J_c'$ is an integer (half-odd integer) if
$M-1$ is even (odd).

Considering the weak coupling limit $c\rightarrow 0^+$, we have
\begin{eqnarray}
 2(I_{j+1}-I_j) &=& \frac{(k_{j+1}-k_j)L }{\pi}+ \sum_{\alpha=1}^M \left[ {\rm
sgn}(k_{j+1}-\lambda_\alpha)-{\rm sgn}(k_j-\lambda_\alpha) \right]
\nonumber
\\
 2(J_{\gamma+1}-J_\gamma +1) &=&  \sum_{l=1}^N \left[ {\rm
sgn}(\lambda_{\gamma+1}-k_l)-{\rm sgn}(\lambda_\gamma -k_l)
\right]   \nonumber \\&&+\sum_{\beta=1}^{M'}\left[ {\rm
sgn}(\lambda_{\gamma+1}- \mu_\beta) - {\rm
sgn}(\lambda_\gamma-\mu_\beta)
\right]     \label{eqcc} \\
2(J_{c+1}'-J_c')  &=&  \sum_{\alpha=1}^M \left[ {\rm
sgn}(\mu_{c+1}-\lambda_\alpha)-{\rm sgn}(\mu_c-\lambda_\alpha)
\right] \nonumber
\end{eqnarray}
If we set $J_{c+1}'-J_c'= m$, there will be $m$ $\lambda_\alpha 's$ satisfying $\mu_c<\lambda_\alpha<\mu_{c+1}$.
Letting $I_{j+1}-I_j = n$, if there is a $\lambda_\alpha$ satisfying $k_j<\lambda_\alpha < k_{j+1}$, we will get
$k_{j+1}-k_j= 2 \pi (n-1)/ L$, otherwise $k_{j+1}-k_j=2 \pi n /L$. From those above, we find that adding a
$\lambda_\alpha$ into the $k_j$ space will expand the distance of neighboring particles and lead to decrease of
the energy. Thus the more $\lambda_\alpha$ lying between $k_j$ and $k_{j+1}$, the lower the energy will be.
Letting $J_{\gamma+1}-J_\gamma = n'$, there will be $n'+1$ solutions of $k_l$ and $\mu_\beta$ satisfying
$\lambda_\gamma<k_l$, $\mu_\beta<\lambda_{\gamma+1}$, where $n'$ is the integer which equals or surpasses one,
i.e. $n'=1$. Then there will be two fermions or two bosons or one fermion and one boson between $\lambda_\gamma$
and $\lambda_{\gamma+1}$. As we always set the distance between the neighboring quantum numbers as one, the most
reasonable state should be one fermion and one boson between $\lambda_\gamma$ and $\lambda_{\gamma+1}$.

Therefore, the state of $M=N, M'=N$ is a boson state which has the
lowest energy. This result concides with parts of Lai's~\cite{Lai}
results. The density of the ground state is the same as
Fig.\ref{Figure Graph1}(left). When $M'=N-1,M=N$, i.e. there is
one spin-up fermion in the ground state, we have the same result
as Fig.\ref{Figure Graph1}(middle). And the more spin up fermions
there are, the higher the energy should be. The particle-hole
excitation spectrum is as Fig.\ref{Figure Graph4}. Due to the
restriction given by the Young tableau, $M=N-2$ gives the
excitation of adding one spin-up and one spin-down fermion, and
exactly there are three holes in the $J$-sequence. This excitation
is depicted in Fig.\ref{Figure Graph5}(b). It also sustains the
analysis of the case $1$.

\subsection{The consistency of the three cases}

In our analytic and numerical results, the density of ground state
(Fig.\ref{Figure Graph1}(left)) and the energy-momentum spectrum
of low-lying excitations (Fig.\ref{Figure Graph5}(a),
Fig.\ref{Figure Graph5}(b)) in the three cases are the same, and
the ground states for all the three cases are the states with
merely bosons. i.e. in the BFF case, the ground state is
$``M=M'=0"$, $``M=N, M'=0"$ for the FBF case and $``M=M'=N"$ for
the FFB case, they are all merely boson state. Thus we conclude
that the properties of this multi-component system in one
dimension are independent of the reference state which we choose,
which manifests the consistency of these cases. Such kind of
equivalence was noticed in the so called super-symmetry t-J
model~\cite{Fabian}.

\section{The ground state phase diagram in the presence of external magnetic fields}\label{sec:phase}

\begin{figure}[h]
\includegraphics[width=0.4 \textwidth,height=4.6cm]{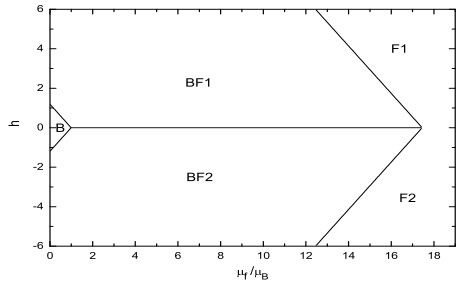}
\caption{\label{Figure Graph8} The phase diagram of particles with
strong coupling limit  $c \rightarrow\infty$.}
\end{figure}

We take into account both the chemical potentials and the external
magnetic fields which bring about the Zeeman splitting for
fermions. At first, we analyze two limit cases: the case of free
particles and the case of particles with infinitely strong
interaction. We have got the phase diagram as Fig.\ref{Figure
Graph7} and Fig.\ref{Figure Graph8}.

\subsection{Weak-coupling limit}

\begin{figure}
\includegraphics[width=0.4 \textwidth,height=6cm]{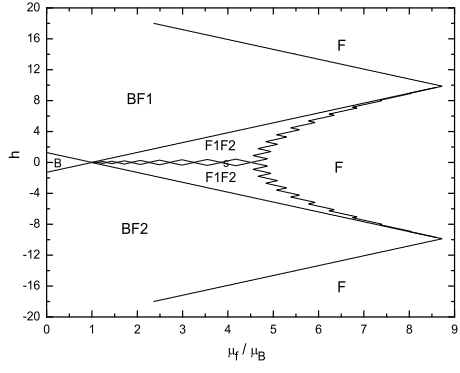}
\caption{\label{Figure Graph7} The phase diagram of free
particles. The vertical axis stands for the external magnetic
field, and the horizontal axis stands for the ratio of chemical
potential between boson ($\mu_b$) and fermion ($\mu_f$). The
region ``B" stands for the state with merely bosons. ``BF1" means
the mix state of bosons and the spin up fermions. ``BF2" means the
mix state of bosons and the spin down fermions. ``F1F2" stands for
the mix state of bosons and two kinds of fermions, but
$N_\uparrow\neq N_\downarrow$. The region ``S" means the spin
singlet state, i.e. $N_\uparrow= N_\downarrow$. The region ``F"
means there is no bosons in ground state.}
\end{figure}

When $c\rightarrow0$, we can treat these particles as free
particles. Thus in the ground state, only fermions have non-zero
momentum because of the pauli excluding principle. We plot the
phase diagram in Fig.\ref{Figure Graph7} for 42 particles.
The region ``S" is composed by some rhombuses that stands for
different number of Fermi pairs$(1,3,5,7,\cdots)$. We take two
rhombuses (3 and 5 pairs of fermions) for example. In the valley
region between them (here we consider h$>$0), there are 5 spin up
and 3 spin down fermions. Thus on the left boundary of this
valley, the energy of 3 pairs of spin singlet fermions equals that
of 5 spin up plus 3 spin down fermions. Because there are $N-M$
bosons, $M-M'$ spin-up fermions, and $M'$ spin-down fermions,
\begin{eqnarray}
H_{3\uparrow3\downarrow}&=&E_{33}-\mu_B\bigl((N-M)+\frac{\mu_f}{\mu_B}M\bigr)-\frac{h}{2}(M-2M^{'})\nonumber\\
                        &=&2{({L\over 2\pi})}^2-\mu_B(36+6\frac{\mu_f}{\mu_B})
\end{eqnarray}
Analogously,
\begin{eqnarray}
H_{5\uparrow3\downarrow}&=&E_{53}-\mu_B\bigl((N-M)+\frac{\mu_f}{\mu_B}M\bigr)-\frac{h}{2}(M-2M^{'})\nonumber\\
                        &=&12{({L\over
                        2\pi})}^2-\mu_B(34+8\frac{\mu_f}{\mu_B})-h
\end{eqnarray}
Thus the phase boundary can be described as
$H_{3\uparrow3\downarrow}=H_{5\uparrow3\downarrow}$, which reduces
to: $ 2\triangle+h=10{({L\over 2\pi})}^2 $ Similarly, the equation
of the right boundary is: $ 2\triangle-h=8{({L\over 2\pi})}^2 $ in
which $\triangle = \mu_f-\mu_B$.

\subsection{Strong-coupling limit}

When the intensity of interaction approaches to
infinity, any two particles can not stay in the same state. Thus
there should be a much stronger excluding force than pauli
principle.
 Because the energy of $N$ particles($N-M$ bosons, $M-M'$ spin up fermions, $M'$ spin down fermions) is:
$
H=E_0-\mu_B\bigl((N-M)+\frac{\mu_f}{\mu_B}M\bigr)-\frac{h}{2}(M-2M'),
$ when the direction of the magnetic field is along the spin
up($h>0$) direction, the spin down fermions can not appear in the
ground state. Therefore, in region ``BF1", there are only fermions
with spin up and bosons; And in region ``F1", there are N fermions
with spin up. The same analysis hold when $h<0$.

\subsection{In general case}

We solve this model within the rules of the Young tableau
(N-M$\geq$ M-$M{'}$ $\geq$ $M{''}$) by using
Eqs.(\ref{eq:BAEBFF}), thus the case of all the particles being
fermions in ground state can not be gained here.
\begin{figure}[h]
\includegraphics[width=0.4 \textwidth,height=4.6cm]{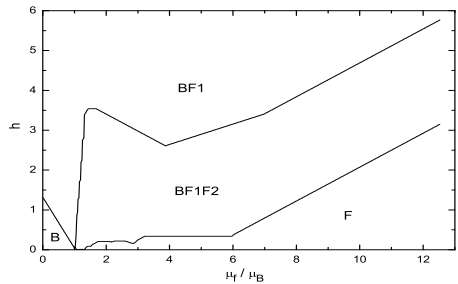}
\caption{\label{Figure Graph9} A sketch of phase when interaction
strength $c=1$}
\end{figure}
And we only computed the case of $h>0$ because of symmetry. As
shown in Fig.\ref{Figure Graph9}, the phase boundary between ``B"
and ``BF1" is the same as that in the two limit cases above, which
intersects with the h-axis at $h=2\mu_B$. The boundary between
``BF1" and ``BF1F2" can be divided into two parts from
$\mu_f/\mu_B\approx1.5$. The left part has some fluctuations which
means adding a pair of spin up fermions in ``BF1F2" side. The
highest point ($\mu_f/\mu_B\approx1.5$) stands for 19 spin up and
1 spin down fermions. According the Young tableau, it can not add
spin up fermion pair on the ``19 up 1 down" state any more. Thus
when $\mu_f/\mu_b>1.5$, we can only add the spin down Fermi pair.
This is the reason why there is a inflexion at
$\mu_f/\mu_B\approx1.5$. We figure out an outline of the spin
singlet phase ``F", in which we can also see that there are some
small fluctuations which stand for adding a pair of spin singlet
fermions. It somewhat likes the free particles case. Because of
pauli excluding principle, only two fermions with different spins
can stay in the same state. Hence in the spin singlet phase, the
pair numbers are $1,3,5\cdots$ along the horizontal axis. From
$\mu_f/\mu_B\approx6$ on, the pair number equals 14, which is the
largest pair number we can get here.


\section{Summary and discussion}

In summery, we have explicitly derived the Bethe-ansatz
equation for the model of one-dimensional Bose-Fermi mixture by means of QISM.
We analyzed the properties of the ground state and the
low-lying excitations on the basis of the Bethe-ansatz
equations. We found that the ground state of this
system is the state with merely bosons.
The low-lying excitations were
discussed extensively.
The energy-momentum spectrum for three types of excitations,
holon-particle, one fermion, two fermions, were plotted for
$c=10.0$ and $c=1.0$. We discussed the phase
diagram of the ground state in the presence of external magnetic
fields and chemical potential,
from which we can know about the populations of boson and fermion
at a given magnetic field and chemical potential.

\appendix

\section{Permutation matrices and the generators of SU(3) Lie algebra}
For BFF case the permutation matrix reads
\begin{eqnarray}\label{eq:permutation_bff}
P_1=\left(\begin{array}{cccccccccccc}1&   &0& &0& &0&0 &0 &0&0 &0 \\
                                   0&   &0& &0& &1&0 &0 &0&0 &0 \\
                                   0&   &0& &0& &0&0 &0 &1&0 &0 \\
                                   0&   &1& &0& &0&0 &0 &0&0 &0 \\
                                   0&   &0& &0& &0&-1&0 &0&0 &0 \\
                                   0&   &0& &0& &0&0 &0 &0&-1&0 \\
                                   0&   &0& &1& &0&0 &0 &0&0 &0 \\
                                   0&   &0& &0& &0&0 &-1&0&0 &0 \\
                                   0&   &0& &0& &0&0 &0 &0&0 &-1\\
   \end{array}\right);
\end{eqnarray}

For FBF case it reads
\begin{eqnarray}\label{eq:permutation_fbf}
P_2=\left(\begin{array}{cccccccccccccc}-1&  0& 0 & 0& &0& &0 &0 &0 &0 \\
                                      0&  0& 0 & 1& &0& &0 &0 &0 &0 \\
                                      0&  0& 0 & 0& &0& &0 &-1&0 &0 \\
                                      0&  1& 0 & 0& &0& &0 &0 &0 &0 \\
                                      0&  0& 0 & 0& &1& &0 &0 &0 &0 \\
                                      0&  0& 0 & 0& &0& &0 &0 &1 &0 \\
                                      0&  0& -1& 0& &0& &0 &0 &0 &0 \\
                                      0&  0& 0 & 0& &0& &1 &0 &0 &0 \\
                                      0&  0& 0 & 0& &0& &0 &0 &0 &-1\\
   \end{array}\right);
\end{eqnarray}

and for FFB case it reads
\begin{eqnarray}\label{eq:permutation_ffb}
P_3=\left(\begin{array}{cccccccccccccc}-1&  0& 0 & 0 &0 &0& &0 & &0& &0 \\
                                      0&  0& 0 & -1&0 &0& &0 & &0& &0 \\
                                      0&  0& 0 & 0 &0 &0& &1 & &0& &0 \\
                                      0& -1& 0 & 0 &0 &0& &0 & &0& &0 \\
                                      0&  0& 0 & 0 &-1&0& &0 & &0& &0 \\
                                      0&  0& 0 & 0 &0 &0& &0 & &1& &0 \\
                                      0&  0& 1 & 0 &0 &0& &0 & &0& &0 \\
                                      0&  0& 0 & 0 &0 &1& &0 & &0& &0 \\
                                      0&  0& 0 & 0 &0 &0& &0 & &0& &1 \\
   \end{array}\right);
\end{eqnarray}

The generators for SU(3) Lie algebra are given by:
$$
H_1  = \left( {\begin{array}{*{20}c}
   1 & 0 & 0  \\
   0 & { - 1} & 0  \\
   0 & 0 & 0  \\
\end{array}} \right),
\quad
H_2  = \left( {\begin{array}{*{20}c}
   0 & 0 & 0  \\
   0 & 1 & 0  \\
   0 & 0 & { - 1}  \\
\end{array}} \right);
$$
$$
E_{\alpha _1 }  = \left( {\begin{array}{*{20}c}
   0 & 1 & 0  \\
   0 & 0 & 0  \\
   0 & 0 & 0  \\
\end{array}} \right),
\quad
E_{\alpha 2}  = \left( {\begin{array}{*{20}c}
   0 & 0 & 0  \\
   0 & 0 & 1  \\
   0 & 0 & 0  \\
\end{array}} \right),
\quad
E_{\alpha _1  + \alpha _2 }  = \left( {\begin{array}{*{20}c}
   0 & 0 & 1  \\
   0 & 0 & 0  \\
   0 & 0 & 0  \\
\end{array}} \right),
$$
$$
E_{ - \alpha _1 }  = \left( {\begin{array}{*{20}c}
   0 & 0 & 0  \\
   1 & 0 & 0  \\
   0 & 0 & 0  \\
\end{array}} \right),
\quad
E_{ - \alpha 2}  = \left( {\begin{array}{*{20}c}
   0 & 0 & 0  \\
   0 & 0 & 0  \\
   0 & 1 & 0  \\
\end{array}} \right),
\quad
E_{ - (\alpha _1  + \alpha _2 )}  = \left( {\begin{array}{*{20}c}
   0 & 0 & 0  \\
   0 & 0 & 0  \\
   1 & 0 & 0  \\
\end{array}} \right)
$$

The work is supported by NSFC Grant No.10225419.


\end{document}